\documentclass[epj,referee]{svjour}
\usepackage{graphics}

\newcommand{\gl}{\stackrel{>}{_<}}
\newcommand{\gel}{\stackrel{\stackrel{\scriptstyle _>}{\scriptstyle _=}}{\scriptstyle _<}}
\newcommand{\DD}{\Delta}
\begin{document}
\titlerunning{
Critical behaviour near the Mott metal-insulator transition}
\title{
Critical behaviour near the metal-insulator transition of a doped Mott insulator}
%\subtitle{Do you have a subtitle?\\ If so, write it here}
\author{Y. \=Ono\inst{1}, R. Bulla\inst{2}, A. C. Hewson\inst{3} \and M. Potthoff\inst{4}% etc
% \thanks is optional - remove next line if not needed
%\thanks{\emph{Present address:} Insert the address here if needed}%
}                     % Do not remove
%
%\offprints{}          % Insert a name or remove this line
%
\institute{Department of Physics, Nagoya University, Furo-cho, Chikusa-ku, Nagoya 464-8602, JAPAN 
\and Theoretische Physik III, Elektronische Korrelationen und Magnetismus, Institut f\"ur Physik, Universit\"at Augsburg,  D-86135 Augsburg,
Germany
\and Department of Mathematics, Imperial College, 180 Queen's Gate, London SW7 2BZ, U.K.
\and Lehrstuhl Festk\"orpertheorie, Institut f\"ur Physik, Humboldt-Universit\"at zu Berlin, Germany
}
\date{Received: date / Revised version: date}
% The correct dates will be entered by Springer
%
\abstract{
We have studied the critical behaviour of a doped Mott insulator near the metal-insulator transition for the infinite-dimensional Hubbard model using a linearized form of dynamical mean-field theory. The discontinuity in the chemical potential in the change from hole to electron doping, for $U$ larger than a critical value $U_c$, has been calculated analytically and is found to be in good agreement with the results of numerical methods. We have also derived analytic expressions for the compressibility, the quasiparticle weight, the double occupancy and the local spin susceptibility near half-filling as functions of the on-site Coulomb interaction and the doping. 
\PACS{
%     {68.35.Rh}{Phase transitions and critical phenomena}   \and
      {71.10.Fd}{Lattice fermion models (Hubbard model, etc.)}   \and
      {71.27.+a}{Strongly correlated electron systems; heavy fermions} \and
      {71.30.+h}{Metal-insulator transitions and other electronic transitions}
     } % end of PACS codes
} %end of abstract
\maketitle
%
%%%%%%%%   Introduction   %%%%%%%%%%%%%%%%%%%%%%%%%%%%%%%%%%%
\section{Introduction}
\label{sec:1}
The Mott-Hubbard metal-insulator transition (MIT) is an important but difficult many-body problem which has been studied extensively \cite{Mott}. Significant progress has been achieved in recent years through the application of dynamical mean-field theory (DMFT) \cite{Georges1} to generic microscopic models, such as the single-band Hubbard model. In this approach the lattice problem is mapped onto an impurity problem where a correlated impurity site is embedded in an effective uncorrelated medium that has to be determined self-consistently. The mapping can be shown to be exact for certain models with local interactions in the limit of infinite dimensions \cite{Metzner}, and constitutes an approximate method for finite-dimensional models. Several techniques have been applied to deal with the effective impurity problem, including iterated perturbation theory \cite{Georges1}, non-crossing approximation \cite{Pruschke}, projective self-consistent method (PSCM) \cite{Fisher}, quantum Monte Carlo (QMC) \cite{Jarrell}, exact diagonalization (ED) \cite{Caffarel} and numerical renormalization group (NRG) \cite{Sakai,Bulla1} techniques. These methods have different strengths and limitations, and apparent discrepancies between some of the results have led to a controversial discussion on the nature of the MIT in the infinite-dimensional Hubbard model at low temperatures \cite{Schlipf,Rozenberg,Noack,Bulla2,Krauth}. Recently, however, most of these differences have been resolved \cite{Bulla3}. 

However, not all aspects of the transition have been investigated thoroughly. In particular, the critical behaviour of the doped Mott insulator has not received so much attention as it is more difficult to solve the model off half-filling using purely numerical methods. Iterated perturbation theory, which provides a reasonable description of the MIT at half-filling \cite{Georges1}, is less suitable for calculations in the doped case since ad hoc modifications of the approach become necessary away from half-filling \cite{KK}. 

Recently, we have shown \cite{Bulla,Ono} that a linearized form of the DMFT provides a simple and attractive technique to obtain approximate but analytical results for the critical regime. Using this approach for the single- and for the two-band Hubbard model, it has been demonstrated that the predictions for the critical coupling of the MIT are in very good agreement with most accurate numerical estimates. 

Here we show that the linearized DMFT can be extended to the non-symmetric case. This allows for a comprehensive analytical investigation of the critical behaviour as a function of doping. In particular, we have calculated the discontinuity in the chemical potential on changing from hole to electron doping. The analytic result is in good agreement with the results of numerical methods, NRG as well as ED. It also agrees well with the result of PSCM \cite{Fisher}. In addition, we have analytically determined the compressibility, the quasiparticle weight, the double occupancy and the local spin susceptibility near half-filling as functions of the on-site Coulomb interaction and the doping. These are difficult to calculate using the numerical methods mentioned above. 

%%%%%%%%   Sec 2   %%%%%%%%%%%%%%%%%%%%%%%%%%%%%%%%%%%
\section{Linearized dynamical mean-field theory}
\label{sec:2}
%%%%%%%%   Sec 2.1 %%%%%%%%%%%%%%%%%%%%%%%%%%%%%%%%%%%
%\subsection{Single-band Hubbard model}
%\label{sec:2.1}

We consider the single-band Hubbard model at zero temperature on the Bethe 
lattice \cite{Bethe} with infinite connectivity $q \mapsto \infty$
\begin{eqnarray} 
H= - \sum_{<i,j>,\sigma} t_{i,j} (c_{i\sigma}^{\dagger} c_{j\sigma} 
      + h.c.) + U \sum_{i} n_{i\uparrow}n_{i\downarrow}.  \label{HUB}
\end{eqnarray} 
The nearest-neighbor hopping is scaled as usual: $t_{i,j}=\frac{t}{\sqrt{q}}$ \cite{Metzner}. For the single-band Hubbard model on the Bethe lattice, the DMFT self-consistency equation is simply given by \cite{Georges1}
\begin{equation} 
{\cal G}_0(z)^{-1} = z+\mu-t^2 G(z), \label{SCE1}
\end{equation} 
where $G(z)$ is the local Green's function of the Hubbard model and 
$
{\cal G}_0(z) = (z-\varepsilon_f- \Delta(z))^{-1}
$
is the non-interacting Green's function of an effective single-impurity Anderson model defined by the hybridization function $\Delta(z)$. The DMFT self-consistency cycle starts with a guess for the hybridization function \cite{DMFT}. The solution of the corresponding impurity model yields the impurity Green's function $G(z)$ to be identified with the on-site Green's function of the lattice model. From the self-consistency equation (\ref{SCE1}) ${\cal G}_0(z)$ and thus a new hybridization function $\Delta(z)$ can be calculated. The cycles have to be repeated until self-consistency is achieved. 

The application of numerical methods for solving the effective impurity problem (see ref. \cite{Bulla2}, for example) have established that at moderate coupling the density of states in the metallic phase is characterized by a three-peak structure consisting of the lower and the upper Hubbard bands and a quasiparticle resonance near the Fermi energy in addition. The quasiparticle peak is more or less isolated from the Hubbard bands. On approaching the Mott-insulating state with increasing interaction strength, the quasiparticle peak becomes extremely narrow and finally vanishes for $U=U_c$.

The linearized dynamical mean-field theory \cite{Bulla,Ono} focuses on this critical regime close to the transition. Here, the quasiparticle peak is approximated by a single pole at the Fermi level, i.\ e.\ 
$
G(z) = \frac{Z}{z}
$ 
near the Fermi level with a small weight $Z \to 0$ as $U \to U_c$. 
Correspondingly, the hybridization function $\Delta(z)$ is a one-pole function
$
\Delta(z)=\frac{V^2}{z} . 
$
This effectively represents an approximate mapping of the model (\ref{HUB}) onto a two-site or zero-bandwidth Anderson model \cite{Hewson} 
\begin{eqnarray} 
H_{\rm 2-site}&=& \varepsilon_f \sum_{\sigma} f^{\dagger}_\sigma f_\sigma +U f^{\dagger}_{\uparrow}f_{\uparrow}f^{\dagger}_{\downarrow}f_{\downarrow} \nonumber \\ 
&+&\varepsilon_c \sum_{\sigma} c^{\dagger}_{\sigma} c_{\sigma} + V \sum_{\sigma}(f^{\dagger}_{\sigma}c_{\sigma}+c^{\dagger}_{\sigma}f_{\sigma})
\label{2SITE}
\end{eqnarray} 
with $\varepsilon_c=0$ and $\varepsilon_f=-\mu$. The hybridization strength $V$ has be determined from the self-consistency equation which takes the simple form\begin{eqnarray} 
t^2 Z = V^2. \label{SCE}
\end{eqnarray} 
For a more detailed discussion and for an extension of the approach to the two-band model see refs.\ \cite{Bulla,Ono}.

Now we calculate the critical value of $U$ for the MIT at half-filling. In this case, the chemical potential is fixed to $\mu=\frac{U}{2}$ due to particle-hole symmetry. Close to the MIT $V$ is small, and $V \to 0$ for $U \to U_c$. The weight $Z$ can be calculated from the impurity spectral function and is given by \cite{Hewson} (see also Appendix \ref{Apex:1}) 
\begin{eqnarray} 
Z = 36 \frac{V^2}{U^2}, \label{W}
\end{eqnarray} 
up to second order in $V$. From eqs. (\ref{SCE},\ref{W}), we obtain the critical value 
\begin{eqnarray} 
U_c = 6t. \label{UC}
\end{eqnarray}
This result is in good agreement with the best numerical estimates of NRG 
($U_c=5.88t$) \cite{Bulla2} and ED ($U_c=5.87t$) \cite{Ono} as well as PSCM 
($U_c=5.84t$) \cite{Georges1}. 

When we solve the self-consistency equation (\ref{SCE}) by iteration, $V^2$ increases exponentially with iteration number for $U<U_c$. In this case, the single-pole approximation for $\Delta(z)$ breaks down. On the other hand, for $U>U_c$, the hybridization strength $V^2$ decreases exponentially to give the 
self-consistent value $V^2 = 0$ corresponding to the insulating solution.

%%%%%%%%   Sec 3   %%%%%%%%%%%%%%%%%%%%%%%%%%%%%%%%%%%
\section{Critical behaviour of the single-band Hubbard model}
\label{sec:3}
%%%%%%%%   Sec 3.1 %%%%%%%%%%%%%%%%%%%%%%%%%%%%%%%%%%%
\subsection{Physical quantities}
\label{sec:3.1}

For more extensive studies of the critical region of the Mott MIT, we need the result for the quasiparticle weight up to fourth order in $V$ (see Appendix \ref{Apex:1}),
\begin{eqnarray} 
Z=V^2F(U,\mu)-V^4G(U,\mu), \label{Z}
\end{eqnarray} 
where 
\begin{eqnarray} 
F(U,\mu)&=&\frac{5}{2\mu^2}+\frac{4}{\mu(U-\mu)}+\frac{5}{2(U-\mu)^2}, 
        \label{F} \\
G(U,\mu)&=&\frac{29}{2\mu^4}+\frac{24}{\mu^3(U-\mu)}+\frac{22}{\mu^2(U-\mu)^2}
              \nonumber \\
        & &    +\frac{24}{\mu(U-\mu)^3}+\frac{29}{2(U-\mu)^4}. \label{G}
\end{eqnarray} 
To determine the doping dependence, we also calculate the impurity occupation number up to second order in $V$ (see Appendix \ref{Apex:1})
\begin{eqnarray} 
n=1+V^2 D(U,\mu), \label{n}
\end{eqnarray} 
where 
\begin{eqnarray} 
D(U,\mu)=\frac{2}{(U-\mu)^2}-\frac{2}{\mu^2}. \label{D}
\end{eqnarray} 
We note that, for $\mu \gel \frac{U}{2}$, $D\gel 0$ and thus $n\gel 1$. 
The double occupancy 
$
d=\langle \hat{n}_\uparrow\hat{n}_\downarrow\rangle
$
is given by (see Appendix \ref{Apex:1}) 
\begin{eqnarray} 
d=\frac{2V^2}{(U-\mu)^2}, \label{d}
\end{eqnarray} 
up to second order in $V$.
We also calculate the static local spin susceptibility, $\chi_0 \equiv \chi_{ii}(0)$, with 
\begin{eqnarray} 
\chi_{ii}(\omega) = \int_{-\infty}^{\infty} {\rm d}t {\rm e}^{{\rm i}\omega t} 
                {\rm i}\theta(t) \langle [\hat{S_i}^-(t),\hat{S_i}^+]\rangle, 
\end{eqnarray} 
where $\hat{S_i}^{\pm}$ are the spin raising and lowering operators at the same site $i$ 
and where we set $(g\mu_B)^2/2=1$. At $T=0$, $\chi_0$ is given by 
\begin{eqnarray} 
\chi_0 =\sum_n \frac{|\langle E_n |\hat{S}^+|E_0 \rangle|^2
                        +|\langle E_n |\hat{S}^-|E_0 \rangle|^2}
                        {E_n-E_0},  \label{chi0}
\end{eqnarray} 
where $|E_n\rangle$ are the eigenstates with the eigenenergies $E_n$, and $|E_0\rangle$ is the ground state. By using the results for the 2-site Anderson model in eq.(\ref{chi0}), we obtain (see Appendix \ref{Apex:1})
\begin{eqnarray} 
\chi_0 = \frac{1}{2V^2\left(\frac{1}{\mu}+\frac{1}{U-\mu} \right)}
                      \Bigl(1+O(V^2) \Bigr). \label{chi}
\end{eqnarray}

We now invoke the self-consistency equation (\ref{SCE}) and substitute $V^2$ from eq.(\ref{SCE}) into eqs.(\ref{Z},\ref{n},\ref{d},\ref{chi}). This yields
\begin{eqnarray} 
Z(U,\mu)&=& \left(F(U,\mu)-\frac{1}{t^2}\right)\frac{1}{t^2G(U,\mu)}, 
      \label{Z2} \\
n(U,\mu)&=&1+Z(U,\mu) t^2D(U,\mu), \label{n2} \\
d(U,\mu)&=&2t^2Z(U,\mu)/(U-\mu)^2, \label{d2} \\
\chi_0(U,\mu)&=&\mu(U-\mu)/\left(2Ut^2Z(U,\mu)\right). \label{chi2} 
\end{eqnarray}

%%%%%%%%   Sec 3.2 %%%%%%%%%%%%%%%%%%%%%%%%%%%%%%%%%%%
\subsection{Critical behaviour at half-filling}
\label{sec:3.2}

First we discuss the critical behaviour near the Mott MIT at half-filling \cite{Bulla}. The chemical potential is fixed to $\mu=\frac{U}{2}$ due to particle-hole symmetry. In fact, eq.(\ref{n}) yields $n(U,\frac{U}{2})=1$ for all $U$ as $D(U,\frac{U}{2})=0$ from eq.(\ref{D}). Substituting $\mu=\frac{U}{2}$ into eqs.(\ref{Z2},\ref{d2},\ref{chi2}), we obtain
\begin{eqnarray} 
 Z(U)&=& \frac{18}{11}\left(1-\frac{U}{U_{c}} \right), \label{Z5} \\
 d(U)&=& \frac{4}{11} \left(1-\frac{U}{U_{c}} \right), \label{d5} \\
 \chi_0(U)&=& \frac{11}{24t} \left(1-\frac{U}{U_{c}} \right)^{-1}, 
       \label{chi5}
\end{eqnarray} 
near $U_{c}$ for $U<U_{c}$. The result for $Z$, eq.(\ref{Z5}), has already been obtained by Bulla and Potthoff \cite{Bulla}. 
Using eq.(\ref{n2}) the compressibility, 
$\kappa = (\frac{\partial n}{\partial \mu})_{_U}$, 
is given by 
\begin{eqnarray} 
\kappa(U)=\left.t^2 Z(U)\frac{\partial D(U,\mu)}{\partial \mu} 
          \right|_{\mu=\frac{U}{2}}, \label{k1}
\end{eqnarray} 
at half-filling for $U<U_{c}$. Substituting eqs.(\ref{D},\ref{Z5}) into eq.(\ref{k1}), we obtain 
\begin{eqnarray} 
 \kappa(U)= \frac{16}{33t}\left(1-\frac{U}{U_{c}} \right), \label{k2} 
\end{eqnarray} 
near $U_{c}$ for $U<U_{c}$.

Due to local charge fluctuations the double occupancy is finite even in the insulating phase with $d(U) \to 0$ only for $U \to \infty$. The linearized DMFT, however, predicts a vanishing double occupancy for $U \to U_c$. This indicates that the approach misses these local fluctuations. We nevertheless believe that the trends in the metallic phase are correct on the mean-field level. This situation may be analogous to the mean-field theory of ferromagnetism in the Heisenberg model, where the fluctuations in the magnetization and short-range order cannot be taken into account. 
We also note that the local spin susceptibility diverges in the limit of the MIT $U\to U_{c}$. Even in this limit, however, the uniform spin susceptibility is finite due to the effect of the super-exchange interaction $J=2t^2/U$ \cite{Georges1}.

%%%%%%%%   Sec 3.3 %%%%%%%%%%%%%%%%%%%%%%%%%%%%%%%%%%%
\subsection{Discontinuity in the chemical potential}
\label{sec:3.3}

\begin{figure*}
\begin{center}
\vspace*{0.8cm}
\resizebox{0.45\textwidth}{!}{
\includegraphics{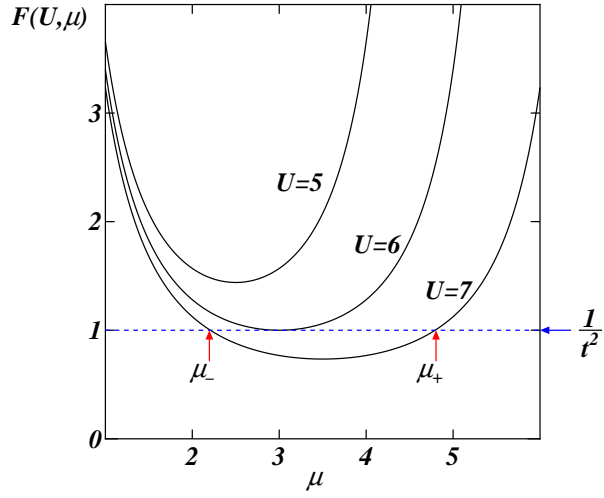}
}
\end{center}
% to insert the figure file. See example above.
% If not, use
%\vspace*{5cm}       % Give the correct figure height in cm
\caption{
The $\mu$ dependence of $F(U,\mu)$ defined in eq.(\ref{F}) for several values of $U$. The critical interaction is $U_{c}=6t$. We set $t=1$. 
}
\label{z0wd}       % Give a unique label
\end{figure*}

For $U<U_c$ the chemical potential $\mu$ as a function of $n$ is continuous at $n=1$. For $U>U_c$, on the other hand, $\mu(n)$ has a discontinuity at $n=1$:
At half-filling the system is a Mott insulator for $\mu_- < \mu < \mu_+$ where 
$\mu_-=\mu_-(U)$, $\mu_+=\mu_+(U)$, and $\mu_-(U) < \mu_+(U)$ for $U>U_c$.
The jump $\Delta \mu = \mu_+ - \mu_-$ can be calculated within the linearized DMFT exploiting the fact that the MIT is characterized by a vanishing quasiparticle weight. Setting $Z=0$ in eq.(\ref{Z2}), we obtain an equation to determine the MIT point:
\begin{eqnarray} 
F(U,\mu)=\frac{1}{t^2}. \label{F2}
\end{eqnarray} 
In Fig.~\ref{z0wd}, $F(U,\mu)$ is plotted as a function of $\mu$ for several values of $U$. When $U<U_{c}$, $F(U,\mu)>\frac{1}{t^2}$ for all $\mu$, and the system is metallic. When $U>U_{c}$, $F(U,\mu)=\frac{1}{t^2}$ for $\mu=\mu_\pm$ ($\mu_+ >\mu_-$), and the system is insulating for $\mu_- <\mu< \mu_+$ while it is metallic for $\mu<\mu_-$ and $\mu>\mu_+$. We notice that, for $U=U_{c}$, $F(U,\mu)$ as a function of $\mu$ is at its minimum for $\mu=\frac{U}{2}$ consistent with the considerations in ref.\ \cite{Ono}.

\begin{figure*}
\begin{center}
\vspace*{0.8cm}
\resizebox{0.45\textwidth}{!}{
\includegraphics{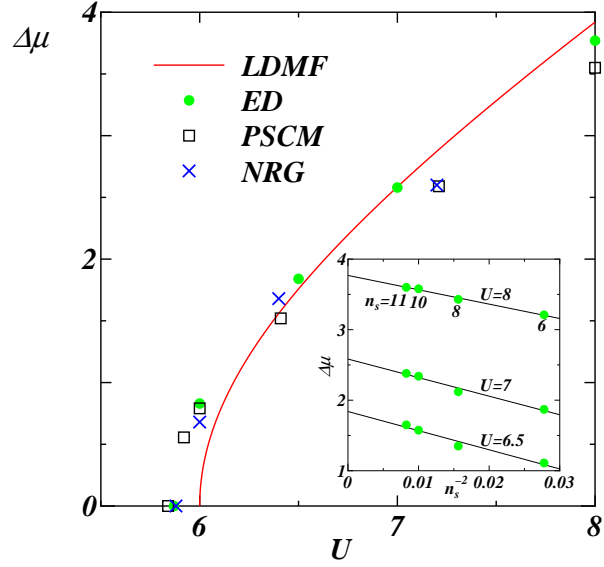}
}
\end{center}
% to insert the figure file. See example above.
% If not, use
%\vspace*{5cm}       % Give the correct figure height in cm
\caption{
The discontinuity of the chemical potential $\Delta\mu=\mu_+-\mu_-$ as a function of $U$ for $U>U_{c}$, as obtained from the linearized DMFT (solid line), from the exact diagonalization method (closed circles), from the projective self-consistent method (open squares) \cite{Fisher} and from the numerical renormalization group method (crosses). Inset shows an extrapolation of the ED results with system sizes $n_s=6, 8, 10$ and $11$. We set $t=1$. 
}
\label{z0egap}       % Give a unique label
\end{figure*}

Using eq.(\ref{F}) we can solve eq.(\ref{F2}) to get the explicit $U$ 
dependence of $\mu_{\pm}$ for $U>U_{c}$ 
\begin{eqnarray} 
\mu_\pm=\frac{U}{2}\pm \frac{U}{2}\left(1+\frac{1}{18u^2}-\sqrt{\frac{10}{9u^2}
        +\left(\frac{1}{18u^2}\right)^2} \right)^{\frac{1}{2}}, \label{mu}
\end{eqnarray} 
where $u \equiv \frac{U}{U_{c}} > 1$. For the discontinuous jump $\Delta\mu = \mu_+ - \mu_-$ we obtain
\begin{eqnarray} 
\Delta\mu= U\left(1+\frac{1}{18u^2}-\sqrt{\frac{10}{9u^2}
        +\left(\frac{1}{18u^2}\right)^2} \right)^{\frac{1}{2}}. \label{Egap}
\end{eqnarray}  
Note that we have $n \to 1$ with $n<1$ ($n>1$) from eq.(\ref{n2}) for 
$\mu\to \mu_-$ ($\mu \to \mu_+$) from the metallic side. 
For $U>U_{c}$ close to $U_{c}$, eq.(\ref{Egap}) yields 
\begin{eqnarray} 
\Delta\mu=\frac{6}{\sqrt{38}}U_{c}\sqrt{\frac{U}{U_{c}}-1}. \label{Egap2}
\end{eqnarray} 
In Fig.~\ref{z0egap} we have plotted $\Delta\mu$, as given by eq.(\ref{Egap}), 
as a function of $U$. 

We have also calculated $\Delta \mu$ numerically by using the NRG and the ED method. 
We solved the full DMFT equation numerically and obtained the occupation number $n$ as a function of $\mu$. When $\mu$ approaches $\mu_+$ ($\mu_-$) from above (below), $n$ approaches unity from above (below) and $n=1$ for $\mu_-<\mu<\mu_+$. We also confirmed that, at $\mu=\mu_{\pm}$, the groundstate changes from singlet ($\mu<\mu_-$ or $\mu>\mu_+$) to doublet ($\mu_-<\mu<\mu_+$). The ED calculations were done for finite cluster sizes $n_s$ up to $n_s=11$. An extrapolation of the results yields the $n_s \to \infty$ extrapolated value of $\Delta\mu$ as shown in the inset of Fig.~\ref{z0egap}.

As can be seen from Fig.~\ref{z0egap}, the result from the linearized DMFT is in good agreement with the results of the NRG and the ED. It also agrees well with the result from the PSCM \cite{Fisher}. We note that the result for $\mu_{\pm}$ from the linearized DMFT seems to be less satisfying for the strong-coupling limit $U\gg U_{c}$. This may be explained by the fact that here $\mu_{\pm}$ is very close to the edge of the lower or upper Hubbard band, respectively, and that the effect of the bandwidth, which is neglected in the linearized DMFT \cite{Bulla}, becomes important \cite{Uc}.

%%%%%%%%%%%%%%%%%% Fig.1 %%%%%%%%%%%%%%%%%%%%%%%%%%%%%
%%%%%%%%%%%%%%%%%%%%%%%%%%%%%%%%%%%%%%%%%%%%%%%%%%%%%%

%%%%%%%%%%%%%%%%%% Fig.2 %%%%%%%%%%%%%%%%%%%%%%%%%%%%%
%%%%%%%%%%%%%%%%%%%%%%%%%%%%%%%%%%%%%%%%%%%%%%%%%%%%%%

In the case of half-filling, as $U$ approaches $U_c$ from below, we have the picture of well separated lower and upper Hubbard bands with a quasiparticle peak in between. The quasiparticle peak disappears at $U=U_c$ leaving an insulator with a finite gap. For $U>U_c$, Fisher {\it et al.} \cite{Fisher} give a similar picture for the slightly doped case in terms of quasiparticle in-gap states. The jump $\Delta \mu$ from hole to electron doping tends to zero as $U \to U_c$ from above, so the Fermi level in the slightly hole-doped case must lie within the quasiparticle peak which in turn lies within the Mott-Hubbard gap. The same applies to the slightly electron-doped case. This interpretation is confirmed by more extensive NRG and ED calculations.

%%%%%%%%   Sec 3.4 %%%%%%%%%%%%%%%%%%%%%%%%%%%%%%%%%%%
\subsection{Doping dependence}
\label{sec:3.4}

Finally, we discuss the critical behaviour of the doped metallic system for $U>U_c$ close to the Mott insulating phase. Substituting eq.(\ref{mu}) into eq.(\ref{n2}), we obtain the doping dependence of the quasiparticle weight for $n \to 1$ and $U>U_{c}$
\begin{eqnarray} 
Z(U,n)=C(U)|n-1|, \label{Z3}
\end{eqnarray} 
where the coefficient $C(U)$ is given by 
\begin{eqnarray} 
C(U)^{-1}=t^2|D(U,\mu_\pm)|=32t^2\frac{U \Delta\mu}{[U^2-(\Delta\mu)^2]^2}.
 \label{C}
\end{eqnarray} 
Due to particle-hole symmetry we obtain the same result for both, $n<1$ and $n>1$. 

Eliminating $Z$ from eqs.(\ref{Z2},\ref{n2}), the occupation number can be expressed as a function of $U$ and $\mu$. For $U>U_{c}$ and close to the MIT we have
\begin{eqnarray} 
n(U,\mu)= 1+\left(F(U,\mu)-\frac{1}{t^2}\right)\frac{D(U,\mu)}{G(U,\mu)}.
      \label{n3}
\end{eqnarray} 
This also yields the compressibility in the limit $n\to 1$ for $U>U_{c}$
\begin{eqnarray} 
 \kappa(U)= \frac{D(U,\mu_\pm)}{G(U,\mu_\pm)} \left.\frac{\partial F(U,\mu)}
               {\partial \mu}\right|_{\mu=\mu_\pm}. \label{kappa}
\end{eqnarray} 
%\begin{eqnarray} 
%F^\prime=-\frac{5}{\mu^3}-\frac{4}{\mu^2(U-\mu)}
%                +\frac{4}{\mu(U-\mu)^2}+\frac{5}{(U-\mu)^3}. 
%\end{eqnarray} 
Again, due to particle-hole symmetry, the result eq.(\ref{kappa}) for $\kappa$ is the same for both, $n<1$ and $n>1$.

Substituting eqs.(\ref{mu},\ref{Z3}) into eq.(\ref{d2}), the double occupancy is obtained as a function of $U$ and $n$ for $n \to 1$ and $U>U_{c}$ 
\begin{eqnarray} 
  d(U,n)=\alpha_\pm(U)|n-1|, \label{d3}
\end{eqnarray} 
where 
\begin{eqnarray} 
 \alpha_\pm(U)^{-1} = \frac{16(U-\mu_\pm)^2 U\Delta\mu}{[U^2-(\Delta\mu)^2]^2},
  \label{alpha}
\end{eqnarray} 
and where $+$ and $-$ stand for $n > 1$ and $n<1$, respectively. For a given interaction strength $U$ we find  $\alpha_+>\alpha_-$. 
This means that when doping the system away from half-filling, the increase of the double occupancy for $n>1$ is stronger than its decrease for $n<1$. 
As mentioned before, there is a finite double occupancy even for $n=1$ and $U>U_c$ in contrast to the result eq.(\ref{d3}) from the linearized DMFT. Nevertheless, we believe that the trend in the metallic state, i.e. the coefficient $\alpha_\pm$, is physically significant.

Substituting eqs.(\ref{mu},\ref{Z3}) into eq.(\ref{chi2}), the local spin susceptibility is obtained as a function of $U$ and $n$ for $n \to 1$ and $U>U_{c}$ 
\begin{eqnarray} 
  \chi_0(U,n)=\beta(U)|n-1|^{-1}, \label{chi3}
\end{eqnarray} 
where 
\begin{eqnarray} 
 \beta(U) = \frac{4\Delta\mu}{U^2-(\Delta\mu)^2}. 
  \label{beta}
\end{eqnarray}

\begin{figure*}
\begin{center}
\vspace*{0.8cm}
\resizebox{0.39\textwidth}{!}{
\includegraphics{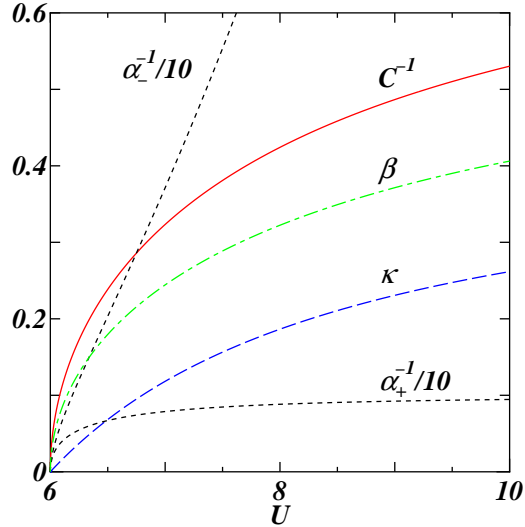}
}
\end{center}
% to insert the figure file. See example above.
% If not, use
%\vspace*{5cm}       % Give the correct figure height in cm
\caption{
The compressibility $\kappa$, the coefficient $C$ of the quasiparticle weight $Z=C|n-1|$, the coefficient $\alpha_\pm$ of the double occupancy $d=\alpha_\pm |n-1|$ with $\pm$ for $n \gl 1$ and the coefficient $\beta$ of the local spin susceptibility $\chi_0=\beta|n-1|^{-1}$ near half-filling as functions of $U$ for $U>U_{c}=6t$ obtained from the linearized DMFT. We set $t=1$. 
}
\label{z0xc}       % Give a unique label
\end{figure*}

In Fig.~\ref{z0xc}, $C^{-1}$, $\kappa$, $\alpha_\pm^{-1}$ and $\beta$, as given by eqs.(\ref{C},\ref{kappa}, \ref{alpha},\ref{beta}), are plotted as functions of $U$. To see the critical properties near $U_{c}$, we substitute eq.(\ref{Egap2}) into eqs.(\ref{C},\ref{kappa},\ref{alpha}, \ref{beta}), and obtain for $U>U_{c}$ 
\begin{eqnarray} 
 C(U)^{-1}&=&\frac{16}{3\sqrt{38}}\sqrt{\frac{U}{U_{c}}-1}, \label{C2} \\
 \kappa(U)&=&\frac{32}{33t}\left(\frac{U}{U_{c}}-1 \right), \label{kappa2} \\
 \alpha_{\pm}(U)^{-1}&=&\frac{24}{\sqrt{38}}\sqrt{\frac{U}{U_{c}}-1}, 
                                                           \label{alpha2}  \\
 \beta(U)&=&\frac{4}{\sqrt{38}t}\sqrt{\frac{U}{U_{c}}-1}.  \label{beta2}
\end{eqnarray} 
The critical properties near $U_{c}$ as resulting from the linearized DMFT are similar to those of the Brinkman-Rice approach \cite{Vollhardt,Vollhardt2}. However, the coefficients are different.

%%%%%%%%%%%%%%%%%% Fig.3 %%%%%%%%%%%%%%%%%%%%%%%%%%%%%
%%%%%%%%%%%%%%%%%%%%%%%%%%%%%%%%%%%%%%%%%%%%%%%%%%%%%%

%%%%%%%%   Sec 4   %%%%%%%%%%%%%%%%%%%%%%%%%%%%%%%%%%%
\section{Conclusion and discussion}
\label{sec:4}

We have shown that analytical calculations of the critical behaviour in the parameter region close to the Mott MIT can be carried out by using a simple linearized version of the DMFT. This approach gives rather reliable estimates for the critical parameters at the transition point. Considering the single-band Hubbard model, we have calculated the discontinuity in the chemical potential $\mu$ for the change from hole to electron doping. The analytical result is in good agreement with the numerical estimates from the PSCM, the NRG and the ED. The results for small hole and electron doping and for $U>U_c$ can be interpreted in terms of quasiparticle in-gap states which lie within the Mott-Hubbard gap. We have also made predictions for the compressibility, the quasiparticle weight, the double occupancy and the local spin susceptibility near half-filling as functions of $U$ and the doping. It would be interesting to see whether these predictions can be confirmed by some of the more comprehensive numerical realizations of DMFT, and whether there is any experimental evidence in support of them \cite{Imada}.

The simplicity of our approach should also enable similar calculations to be made for more general models, such as multi-band Hubbard-type models, where we can have both Mott-Hubbard type and charge-transfer type metal-insulator transitions \cite{Ono1}. Even for more complex multi-band models one might obtain analytical expressions as has been demonstrated recently for the five critical parameters in the two-band Hubbard model \cite{Ono}. This makes the linearized DMFT a powerful technique for exploring general trends and phase diagrams within a high-dimensional parameter space. 
An interesting question to examine would be the role of in-gap states in the two-band Hubbard model (equivalent to the $d$-$p$ model) near the MIT. In this parameter regime the results of the model should  be relevant for a description of the behaviour of weakly doped high-$T_c$ materials.

%%%%%%%%  Acknowledgments  %%%%%%%%%%%%%%%%%%%%%%%%%%%%
\section{Acknowledgments}

We gratefully acknowledge the support of the Grant-in-Aid for Scientific Research from the Ministry of Education, Science, Sports and Culture, and also by CREST (Core Research for Evolutional Science and Technology) of Japan Science and Technology Corporation (JST) (Y\=O); the Deutsche Forschungsgemeinschaft, through the Sonderforschungsbereich 484 (RB) and 290 (MP); and the EPSRC, through research grant (GR/J85349) (ACH).
We also wish to thank the Newton Institute in Cambridge for providing facilities and support for this work during their six month programme on `Strongly Correlated Electron Systems'.

%%%%%%%%  Appendix  %%%%%%%%%%%%%%%%%%%%%%%%%%%%%%%%%%%
\appendix
\section{\hspace{-1mm}ppendix A}
\label{Apex:1}

Here we discuss the two-site Anderson model eq.(\ref{2SITE}) \cite{Hewson} in the limit 
$V\to 0$. 
We assume that the conduction level is between the atomic $f$-level and the upper Hubbard level 
$
\varepsilon_f<\varepsilon_c < \varepsilon_f+U. 
$
Then we define 
\begin{eqnarray} 
 \DD &\equiv& \varepsilon_c-\varepsilon_f>0, \\
 W &\equiv& \varepsilon_f+U -\varepsilon_c>0. 
\end{eqnarray} 

The one-electron eigenstates 
\begin{eqnarray} 
|E_{\pm}\rangle =  a_{\pm}f_\sigma^+|0\rangle 
                 + b_{\pm}c_\sigma^+|0\rangle ,
\end{eqnarray} 
correspond to the eigenenergies
\begin{eqnarray} 
E_{\pm} =  \frac{1}{2}\left(\varepsilon_c+\varepsilon_f
                  \pm\sqrt{(\varepsilon_c-\varepsilon_f)^2+4V^2}\right). 
         \label{EPM}
\end{eqnarray} 
For small hybridization strength
$V^2 \ll \Delta$, 
eq.(\ref{EPM}) is simplified as
\begin{eqnarray} 
E_{+} &=&  \varepsilon_c +\frac{V^2}{\DD}-\frac{V^4}{\DD^3}, \\ 
E_{-} &=&  \varepsilon_f -\frac{V^2}{\DD}+\frac{V^4}{\DD^3}, 
\end{eqnarray}
to fourth order in $V$, with the corresponding eigenstates 
\begin{eqnarray} 
|E_{+}\rangle =  \alpha\left\{\frac{V}{\DD}\left(1-\frac{V^2}{\DD^2} \right) 
                 f_\sigma^+ + c_\sigma^+  \right\} |0\rangle,       \\
|E_{-}\rangle =  \alpha\left\{f_\sigma^+ - \frac{V}{\DD} 
          \left(1-\frac{V^2}{\DD^2} \right)  c_\sigma^+  \right\} |0\rangle,
\end{eqnarray} 
where
\begin{eqnarray} 
\alpha^2 = 1-\frac{V^2}{\DD^2} +3\frac{V^4}{\DD^4}. 
\end{eqnarray} 
Similarly, we obtain the three-electron (one-hole) eigenenergies 
\begin{eqnarray} 
\bar{E}_{\pm} =  \frac{1}{2}\left(3\varepsilon_c+3\varepsilon_f+U
                \pm\sqrt{(\varepsilon_f+U-\varepsilon_c)^2+4V^2}\right). 
         \label{EPMB}
\end{eqnarray} 
For small hybridization strength
$V^2 \ll W$, 
eq.(\ref{EPMB}) is simplified as
\begin{eqnarray} 
\bar{E}_{+} &=&  \varepsilon_c +2\varepsilon_f+U 
               +\frac{V^2}{W}-\frac{V^4}{W^3}, \\
\bar{E}_{-} &=&  2\varepsilon_c+\varepsilon_f 
               -\frac{V^2}{W}+\frac{V^4}{W^3},
\end{eqnarray}
to fourth order in $V$. The corresponding eigenstates are 
\begin{eqnarray} 
|\bar{E}_{+}\rangle &=&  \bar{\alpha}\left\{-\frac{V}{W}\left(1-\frac{V^2}{W^2} 
                 \right)  f_\sigma + c_\sigma  \right\} |4\rangle,     \\
|\bar{E}_{-}\rangle &=&  \bar{\alpha}\left\{f_\sigma + \frac{V}{W} 
          \left(1-\frac{V^2}{W^2} \right)  c_\sigma  \right\} |4\rangle,
\end{eqnarray} 
where 
\begin{eqnarray} 
\bar{\alpha}^2 = 1-\frac{V^2}{W^2} +3\frac{V^4}{W^4}, 
\end{eqnarray} 
and 
$
|4\rangle=f_\uparrow^+ f_\downarrow^+ c_\uparrow^+ c_\downarrow^+|0\rangle.
$

The two electron states can be classified as singlets or triplets. In the triplet state, the spatial part of the wavefunction is antisymmetric and the interaction $U$ has no effect. Then the total energy of the triplet state is given by 
$
E_{+} + E_{-} = \varepsilon_c+\varepsilon_f. 
$
There are three possible singlet states which can be written by the linear combination of the states,
\begin{eqnarray} 
|\phi_1\rangle &=& \frac{1}{\sqrt{2}} (c_\uparrow^+ f_\downarrow^+ 
                      - c_\downarrow^+ f_\uparrow^+ )|0\rangle,  \\
|\phi_2\rangle &=& c_\uparrow^+ c_\downarrow^+ |0\rangle,  \\
|\phi_3\rangle &=& f_\uparrow^+ f_\downarrow^+ |0\rangle. 
\end{eqnarray} 
The eigenenergies are given by the solutions of the equation,
\begin{eqnarray} 
\left|
\begin{array}{ccc}
E-\varepsilon_c-\varepsilon_f \  & -\sqrt{2}V & -\sqrt{2}V \\
 -\sqrt{2}V & E-2\varepsilon_c & 0 \\
 -\sqrt{2}V & 0 & E-2\varepsilon_f -U \\
\end{array}
\right|
=0.
\end{eqnarray}
To fourth order in $V$, the ground state eigenenergy is 
\begin{eqnarray} 
E_{0} =    \varepsilon_c +\varepsilon_f  -2V^2
               \left(\frac{1}{\DD}+\frac{1}{W} \right)
               \left(1-\frac{2V^2}{\DD^2}-\frac{2V^2}{W^2} \right), 
\end{eqnarray}
and the corresponding singlet ground state is 
\begin{eqnarray} 
|E_{0}\rangle = \alpha_0\left\{|\phi_1\rangle 
           -\frac{\sqrt{2}V}{\DD} \left(1-\frac{2V^2}{\DD^2}
                     -\frac{2V^2}{\DD W} \right)  |\phi_2\rangle \right. 
                     \nonumber \\
      \left.-\frac{\sqrt{2}V}{W}  \left(1-\frac{2V^2}{W^2}
                      -\frac{2V^2}{\DD W} \right)  |\phi_3\rangle
            \right\} , \label{E0}
\end{eqnarray}
with 
\begin{eqnarray} 
\alpha_0^2 &=& 1-2V^2\left(\frac{1}{\DD^2}+\frac{1}{W^2} \right) \nonumber \\
           &+&4V^4\left(\frac{3}{\DD^4}+\frac{2}{\DD^3 W}+\frac{2}{\DD^2 W^2}
                      +\frac{2}{\DD W^3}+\frac{3}{W^4}\right).
\end{eqnarray}

Now we calculate the $f$-electron Green's function of this model. When a $f,\uparrow$ electron is removed from the ground state $|E_{0}\rangle$, there are two possible final states: $|E_{+}\rangle$ and $|E_{-}\rangle$. Correspondingly, there are two possible single-hole excitations with excitation energies, 
\begin{eqnarray} 
E_+-E_0 &=& -\varepsilon_f+V^2\left(\frac{3}{\DD}+\frac{2}{W} \right) 
               \nonumber \\
        &-&V^4\left(\frac{5}{\DD^3}+\frac{4}{\DD^2 W}+\frac{4}{\DD W^2}
                                   +\frac{4}{W^3}\right) \nonumber \\
        &\equiv& -\varepsilon_1, \label{E1} \\
E_--E_0 &=& -\varepsilon_c+V^2\left(\frac{1}{\DD}+\frac{2}{W} \right) 
               \nonumber \\
        &-&V^4\left(\frac{3}{\DD^3}+\frac{4}{\DD^2 W}+\frac{4}{\DD W^2}
                                   +\frac{4}{W^3}\right) \nonumber \\
        &\equiv& -\varepsilon_2, \label{E2} 
\end{eqnarray}
to fourth order in $V$. The matrix elements for these transitions are
\begin{eqnarray} 
\langle E_+|f_\uparrow|E_0\rangle &=& \frac{\alpha \alpha_0}{\sqrt{2}} 
            \biggl\{1 - \frac{2V^2}{\DD W}   \nonumber \\
             & &\times \left(1-\frac{V^2}{\DD^2} \right)
            \left(1-\frac{2V^2}{W^2}-\frac{2V^2}{\DD W} \right)
            \biggr\}, \nonumber \\
\langle E_-|f_\uparrow|E_0\rangle &=& \frac{\alpha \alpha_0}{\sqrt{2}} 
           \biggl\{-\frac{V}{\DD}\left(1-\frac{V^2}{\DD^2} \right) \nonumber \\
             &-& \frac{2V}{W}\left(1-\frac{2V^2}{W^2}-\frac{2V^2}{\DD W}
             \right) \biggr\}, \nonumber
\end{eqnarray}
which yield the transition probabilities: 
\begin{eqnarray} 
|\langle E_+|f_\uparrow|E_0\rangle|^2 &=& \frac{1}{2} 
             -\frac{V^2}{2}\left(\frac{3}{\DD^2}+\frac{4}{\DD W}+\frac{2}{W^2}
             \right) + \frac{V^4}{2}  \nonumber \\
             &\times& \left(\frac{17}{\DD^4}+\frac{24}{\DD^3 W}
             +\frac{22}{\DD^2 W^2}+\frac{24}{\DD W^3}+\frac{12}{W^4}\right)  
             \nonumber \\
             &\equiv& w_1,  \\
|\langle E_-|f_\uparrow|E_0\rangle|^2 &=& 
             \frac{V^2}{2}\left(\frac{1}{\DD^2}+\frac{4}{\DD W}+\frac{4}{W^2}
             \right) -\frac{V^4}{2} \nonumber \\
             &\times& \left(\frac{5}{\DD^4}+\frac{16}{\DD^3 W}
             +\frac{22}{\DD^2 W^2}+\frac{32}{\DD W^3}+\frac{24}{W^4}\right)  
             \nonumber \\
             &\equiv& w_2, 
\end{eqnarray}
to fourth order in $V$. 

When a $f,\uparrow$ electron is added to the ground state $|E_{1}\rangle$, possible final states are $|\bar{E}_{-}\rangle$ and $|\bar{E}_{+}\rangle$. Correspondingly, there are two possible single-particle excitations with excitation energies, 
\begin{eqnarray} 
\bar{E}_--E_0 &=& \varepsilon_c+V^2\left(\frac{2}{\DD}+\frac{1}{W} \right) 
               \nonumber \\
        &-&V^4\left(\frac{4}{\DD^3}+\frac{4}{\DD^2 W}+\frac{4}{\DD W^2}
                                   +\frac{3}{W^3}\right) \nonumber \\
        &\equiv& \varepsilon_3, \label{E3} \\
\bar{E}_+-E_0 &=& \varepsilon_f+U+V^2\left(\frac{2}{\DD}+\frac{3}{W} \right) 
               \nonumber \\
        &-&V^4\left(\frac{4}{\DD^3}+\frac{4}{\DD^2 W}+\frac{4}{\DD W^2}
                                   +\frac{5}{W^3}\right) \nonumber \\
        &\equiv& \varepsilon_4, \label{E4} 
\end{eqnarray}
to fourth order in $V$. The matrix elements for these transitions are
\begin{eqnarray} 
\langle \bar{E}_-|f_\uparrow|E_0\rangle &=& \frac{\bar{\alpha} \alpha_0}
             {\sqrt{2}} \biggl\{\frac{V}{W}\left(1-\frac{V^2}{W^2} \right)                   \nonumber \\
             &+& \frac{2V}{\DD}\left(1-\frac{2V^2}{\DD^2}-\frac{2V^2}{\DD W}
             \right) \biggr\}, \nonumber \\
\langle \bar{E}_+|f_\uparrow|E_0\rangle &=& \frac{\bar{\alpha} \alpha_0}
            {\sqrt{2}} \biggl\{1 - \frac{2V^2}{\DD W}   \nonumber \\
             & &\times \left(1-\frac{V^2}{W^2} \right)
            \left(1-\frac{2V^2}{\DD^2}-\frac{2V^2}{\DD W} \right)
            \biggr\}, \nonumber 
\end{eqnarray}
which yield the transition probabilities: 
\begin{eqnarray} 
|\langle \bar{E}_-|f_\uparrow|E_0\rangle|^2 &=& 
             \frac{V^2}{2}\left(\frac{4}{\DD^2}+\frac{4}{\DD W}+\frac{1}{W^2}
             \right) -\frac{V^4}{2} \nonumber \\
             &\times& \left(\frac{24}{\DD^4}+\frac{32}{\DD^3 W}
             +\frac{22}{\DD^2 W^2}+\frac{16}{\DD W^3}+\frac{5}{W^4}\right)  
             \nonumber \\
             &\equiv& w_3, \\
|\langle \bar{E}_+|f_\uparrow|E_0\rangle|^2 &=& \frac{1}{2} 
             -\frac{V^2}{2}\left(\frac{2}{\DD^2}+\frac{4}{\DD W}+\frac{3}{W^2}
             \right) + \frac{V^4}{2}  \nonumber \\
             &\times& \left(\frac{12}{\DD^4}+\frac{24}{\DD^3 W}
             +\frac{22}{\DD^2 W^2}+\frac{24}{\DD W^3}+\frac{17}{W^4}\right)  
             \nonumber \\
             &\equiv& w_4,  \label{W4}
\end{eqnarray}
to fourth order in $V$. 

From eqs.(\ref{E1}-\ref{W4}), we obtain the $f$-electron Green's function which has four poles:
\begin{eqnarray} 
G_\sigma(z) = \sum_{i=1}^4 \frac{w_i}{z-\varepsilon_i}. 
\end{eqnarray}
In the limit $V\to 0$, high-energy poles at $\varepsilon_1 \approx \varepsilon_f$ and $\varepsilon_4 \approx \varepsilon_f+U$ have large residues $w_1 \approx w_2 \approx \frac12$, while low-energy poles merge together at $\varepsilon_2 \approx \varepsilon_3 \approx 0$ with small total weight $Z \equiv w_2+w_3$: 
\begin{eqnarray} 
  Z &=&   V^2 \left(\frac{5}{2\DD^2}+\frac{4}{\DD W}+\frac{5}{2W^2}
          \right)  \nonumber \\
          &-& V^4 \left(\frac{29}{2\DD^4}+\frac{24}{\DD^3 W}
          +\frac{22}{\DD^2 W^2}+\frac{24}{\DD W^3}+\frac{29}{2W^4}\right),  
          \nonumber 
\end{eqnarray}
to fourth order in $V$. The number of $f$ electrons, $n=2(w_1+w_2)$, is given by
\begin{eqnarray} 
  n = 1   &-&2V^2 \left(\frac{1}{\DD^2}-\frac{1}{W^2}
          \right)  \nonumber \\
          &+& 4V^4 \left(\frac{3}{\DD^4}+\frac{2}{\DD^3 W}
          -\frac{2}{\DD W^3}-\frac{3}{W^4}\right),  
          \nonumber 
\end{eqnarray}
to fourth order in $V$. 
By using eq.(\ref{E0}), we obtain the double occupancy, 
$
d=\langle E_{0}| \hat{n}_{f\uparrow}\hat{n}_{f\downarrow}|E_{0}\rangle, 
$
to fourth order in $V$ 
\begin{eqnarray} 
  d =     \frac{2V^2}{W^2} \left\{1-V^2\left(
          \frac{2}{\DD^2}+\frac{4}{\DD W}+\frac{6}{W^2}
          \right)\right\}.      \nonumber 
\end{eqnarray}

%%%%%%%%  References %%%%%%%%%%%%%%%%%%%%%%%%%%%%%%%%%%%


\begin{thebibliography}{}
\bibitem{Mott}
F. Gebhard, {\em The Mott Metal-Insulator Transition}, Springer Tracts in Modern Physics,
Vol. 137 (Springer, Berlin, 1994).

\bibitem{Georges1}
For a review, see
A. Georges, G. Kotliar, W. Krauth, M. J. Rozenberg, Rev. Mod. Phys.
\textbf{68}, 13 (1996).

\bibitem{Metzner}
W. Metzner, D. Vollhardt,  Phys. Rev. Lett. \textbf{62}, 324 (1989).

\bibitem{Pruschke}
Th. Pruschke, M. Jarrell, J.K. Freericks, Adv. Phys. \textbf{44}, 187 (1995).

\bibitem{Fisher}
D. S. Fisher, G. Kotliar, G. Moeller, Phys. Rev. \textbf{B52}, 17112 (1995). 

\bibitem{Jarrell}
M. Jarrell, Phys. Rev. Lett. \textbf{69}, 168 (1992).

\bibitem{Caffarel}
M. Caffarel, W. Krauth, Phys. Rev. Lett. \textbf{72}, 1545 (1994). 

\bibitem{Sakai}
O. Sakai, Y. Kuramoto, Solid State Commun. \textbf{89}, 307 (1994).

\bibitem{Bulla1}
R. Bulla, A.C. Hewson, Th. Pruschke, J. Phys. Cond. Mat. \textbf{10}, 8365 (1998).

\bibitem{Schlipf}
J. Schlipf,  M. Jarrell, P.G.J. van Dongen, N. Bl\"umer, S. Kehrein, Th. Pruschke, D. Vollhardt, Phys. Rev. Lett. \textbf{82}, 4890 (1999).

\bibitem{Rozenberg}
M. J. Rozenberg, R. Chitra, G. Kotliar, Phys. Rev. Lett. \textbf{83}, 3498 (1999). 

\bibitem{Noack}
R. M. Noack, F. Gebhard, Phys. Rev. Lett. \textbf{82}, 1915 (1999). 

\bibitem{Bulla2}
R. Bulla, Phys. Rev. Lett. \textbf{83}, 136 (1999). 

\bibitem{Krauth}
W. Krauth,  Phys. Rev. B \textbf{62}, 6860 (2000). 

%\bibitem{Joo}
%J. Joo, V. Oudovenko: cond-mat/0009367.

\bibitem{Bulla3}
R. Bulla, T.A. Costi, D. Vollhardt, cond-mat/0012329.

\bibitem{KK}
H. Kajueter, G. Kotliar, Phys. Rev. Lett. \textbf{77}, 131 (1996);
M. Potthoff, T. Wegner, W. Nolting, Phys. Rev. B \textbf{55}, 16132 (1997).

\bibitem{Bulla}
R. Bulla, M. Potthoff, Eur. Phys. J. B \textbf{13}, 257 (2000).

\bibitem{Ono}
Y. \=Ono, R. Bulla, A. C. Hewson, Eur. Phys. J. B \textbf{19}, 375 (2001). 

\bibitem{Bethe}
This simplification is not essential. The general case for the single-band Hubbard model and for the two-band Hubbard model are discussed in refs. \cite{Bulla,Ono}.

\bibitem{DMFT}
One could start with a guess for the self-energy as well. 

\bibitem{Hewson}
A. C. Hewson, \textit{The Kondo Problem to Heavy Fermions} (Cambridge Univ. Press, 1993).

\bibitem{Uc}
In the limit $U =\infty$, eq.(\ref{mu}) yields $\mu_- = \sqrt{\frac{5}{2}}t$. 
This underestimates the numerical value obtained from the PSCM $\mu_- \approx 2.036t$ 
\cite{Fisher}. The latter is consistent with a quasiparticle peak separated from the lower 
Hubbard band with half-bandwidth $2t$ for $U =\infty$. 

\bibitem{Vollhardt}
D. Vollhardt, Rev. Mod. Phys. \textbf{56}, 99 (1984). 

\bibitem{Vollhardt2}
D. Vollhardt, P. W\"olfle, P. W. Anderson, Phys. Rev. B \textbf{35}, 6703 (1987). 

\bibitem{Imada}
M. Imada, A. Fujimori, Y. Tokura, Rev. Mod. Phys. \textbf{70}, 1039 (1998). 

\bibitem{Ono1}
Y. \=Ono, K. Sano, Proceedings of CREST International Workshop: J. Phys. Chem. Solids \textbf{62}, 285 (2001).

\end{thebibliography}
\end{document}